\documentclass[sigconf]{acmart}
\AtBeginDocument{%
  }

\copyrightyear{2026}
\acmYear{2026}
\setcopyright{cc}
\setcctype{by-nc-nd}
\acmConference[SIGCSE TS 2026]{Proceedings of the 57th ACM Technical Symposium on Computer Science Education V.1}{February 18--21, 2026}{St. Louis, MO, USA}
\acmBooktitle{Proceedings of the 57th ACM Technical Symposium on Computer Science Education V.1 (SIGCSE TS 2026), February 18--21, 2026, St. Louis, MO, USA}
\acmPrice{}
\acmDOI{10.1145/3770762.3772549}
\acmISBN{979-8-4007-2256-1/26/02}

\usepackage{placeins}
\usepackage{comment}
\usepackage{soul}
\usepackage{array}
\usepackage{tabularx}
\usepackage{multirow}
\usepackage{graphicx}
\usepackage{xcolor}
\usepackage{colortbl}
\usepackage[capitalize]{cleveref}
\usepackage{balance}

\usepackage[skins]{tcolorbox}

\definecolor{blueFill}{RGB}{218, 232, 252}
\definecolor{blueBorder}{RGB}{108, 142, 191}
\definecolor{orangeFill}{RGB}{251, 231, 207}
\definecolor{orangeBorder}{RGB}{206, 158, 53}
\definecolor{redFill}{RGB}{241, 208, 205}
\definecolor{redBorder}{RGB}{161, 85, 79}
\definecolor{purpleFill}{RGB}{222, 212, 229}
\definecolor{purpleBorder}{RGB}{142, 113, 159}

\newtcbox{\tagBlue}{enhanced,nobeforeafter,tcbox raise base,boxrule=0.6pt,top=0mm,bottom=0mm,
  right=0mm,left=0mm,arc=0pt,boxsep=1.5pt, colframe=blueBorder,coltext=black,colback=blueFill,
}
\newtcbox{\tagOrange}{enhanced,nobeforeafter,tcbox raise base,boxrule=0.6pt,top=0mm,bottom=0mm,
  right=0mm,left=0mm,arc=0pt,boxsep=1.5pt, colframe=orangeBorder,coltext=black,colback=orangeFill,
}
\newtcbox{\tagRed}{enhanced,nobeforeafter,tcbox raise base,boxrule=0.6pt,top=0mm,bottom=0mm,
  right=0mm,left=0mm,arc=0pt,boxsep=1.5pt, colframe=redBorder,coltext=black,colback=redFill,
}
\newtcbox{\tagPurple}{enhanced,nobeforeafter,tcbox raise base,boxrule=0.6pt,top=0mm,bottom=0mm,
  right=0mm,left=0mm,arc=0pt,boxsep=1.5pt, colframe=purpleBorder,coltext=black,colback=purpleFill,
}

\begin{document}

\title[Teaching Probabilistic ML to Empower Socially \emph{and} Mathematically Informed AI Discourse]{Teaching Probabilistic Machine Learning in the Liberal Arts: Empowering Socially \emph{and} Mathematically Informed AI Discourse}


\author{Yaniv Yacoby}
\affiliation{%
  \institution{Wellesley College}
  \city{Wellesley, MA}
  \country{USA}
  }
\email{yy109@wellesley.edu}

\renewcommand{\shortauthors}{Yaniv Yacoby}

\begin{abstract} 
We present a new undergraduate ML course at our institution, a small liberal arts college serving students minoritized in STEM, designed to empower students to critically connect the mathematical foundations of ML with its sociotechnical implications. 
We propose a ``framework-focused'' approach, teaching students the language and formalism of probabilistic modeling while leveraging probabilistic programming to lower mathematical barriers. 
We introduce methodological concepts through a whimsical, yet realistic theme, the ``Intergalactic Hypothetical Hospital,'' to make the content both relevant and accessible. 
Finally, we pair each technical innovation with counter-narratives that challenge its value using real, open-ended case-studies to cultivate dialectical thinking.
By encouraging creativity in modeling and highlighting unresolved ethical challenges, we help students recognize the value and need of their unique perspectives, empowering them to participate confidently in AI discourse as technologists and critical citizens.
\end{abstract}

\ccsdesc[500]{Social and professional topics~Computing education}
\ccsdesc[500]{Computing methodologies~Machine learning algorithms}
\ccsdesc[500]{Mathematics of computing~Probabilistic representations}
\ccsdesc[500]{Mathematics of computing~Probabilistic inference problems}

\keywords{machine learning, artificial intelligence, data science, ethics, liberal arts, CS education}


\maketitle

\section{Introduction}

Over the past several decades, Artificial Intelligence (AI) and Machine Learning (ML) have become ubiquitous in many high-stakes applications, in which they have caused significant, well-documented, systemic harms.
As an example, consider their use in the U.S.~criminal justice system for pretrial
detention decisions (e.g.~whether to release or detain a defendant before trial).
A few years ago, estimates indicated that about two-thirds of all Americans live in jurisdictions that rely on such tools~\cite{lattimore2020prevalence,copp2022pretrial}.
In this setting, AI has been found to have a higher false-positive rate for Black defendants than for white defendants~\cite{propublica2016,angwin2022machine}.
This bias in pretrial detention compounds as individuals move through the system, from sentencing to prison misconduct predictions to parole decisions (e.g.~\cite{o2021compounding,alexander2021new}), continuing to reinforce systemic racism~\cite{yeung2023part1}.

As projections of economic benefits~\cite{mckinsey2023,goldman2023} drive proliferation of increasingly complex models in safety-critical areas (e.g.~generative AI for therapy~\cite{stade2024large} and drug safety~\cite{hakim2024need}), the potential for harm continues to increase.
While users, affected communities, and researchers uncover new types of harm caused by AI, many are already well-known and can be anticipated (e.g.~\cite{shelby2023sociotechnical}).
\ul{In spite of this, the vast majority of undergraduate courses on the AI spectrum---data science, AI, statistics, and ML---\emph{omit ethics}}~\cite{garrett2020more,oliver2021undergraduate,baumer2022integrating,suarez2022ethical}.

To empower students with critical frameworks to anticipate and address systemic harms caused by AI, we argue that curricula serving courses on the AI spectrum must address three ``layers of bias,'' characterized by~\citet{eckhouse2019layers}.
At the top-layer, we have biases within ML methods and choice of evaluation metrics.
In our running example, matching accuracy across racial groups (instead of false positive rates) presents a flawed notion of fairness~\cite{propublica2016}.
At the middle-layer, we have biases in data quality and collection.
Continuing with our example, re-arrest data is conflated with risk, as opposed to reflecting biased policing practices~\cite{barabas2020beyond}. 
Finally, at the base-layer, we have philosophical (e.g.~moral, legal) concerns about ethics.
As many have argued, the use of AI in pretrial risk assessment is unconstitutional (e.g.~\cite{starr2014evidence,yeung2023part2}).

We observe that existing approaches to AI pedagogy are well-positioned to connect technical content to the ethical questions at the top two layers, but are limited in addressing the base-layer ethics.
This is because base-layer challenges are often solely addressed by non-technical ``AI and society'' courses, overlooking important connections with theoretical content. 
For example, conventional wisdom says that large neural networks (NNs) can learn any pattern from data without human influence---that NNs ``let the data speak for itself.''
In reality, however, recent ML theory connects large NNs to classical kernel regression~\cite{jacot2018neural}, wherein a human-selected ``kernel'' determines how training data is combined for prediction. 
This theory shows that, although obscured by complexity, large NNs remain subject to (implicit) human choices.
Linking theory to ethics, we can challenge the myth of objectivity in NNs and alert students to its dangers by recalling how similar appeals were repeatedly misused to justify scientific racism in the Eugenics movement~\cite{gould1996mismeasure,kennedy2024teaching}.
For example, \citet{pearson1925problem} invoked ``the cold light of statistical inquiry'' to justify claims of Jewish immigrant girls' intellectual inferiority when attempting to shape immigration policy.



In this work, we describe a curricular initiative at our institution---a small, liberal arts college serving students with minoritized backgrounds in STEM---to connect technical content to base-layer ethics.
Our course is an elective for upper-level undergrads with little to no prior ML experience.
In contrast to existing approaches to AI pedagogy, we propose a ``framework-focused'' lens, teaching students the language and formalism of probabilistic modeling while leveraging probabilistic programming to lower mathematical barriers. 
We introduce methodological concepts through a whimsical, yet realistic theme, the ``Intergalactic Hypothetical Hospital,'' to make the content relevant and accessible. 
Finally, we pair each technical innovation with counter-narratives that challenge its value using real, open-ended case-studies to cultivate dialectical thinking.
By encouraging creativity in modeling and highlighting unresolved ethical challenges, we help students recognize the value and need of their unique perspectives, empowering them to participate confidently in AI discourse as technologists and critical citizens.

We evaluated the course via an anonymous survey, asking students how they responded to key pedagogical design choices.
The survey provides qualitative evidence of the students' experiences, suggesting they felt empowered to participate in AI discourse as technologists and critical citizens.
We also provide an analysis of the limitations of our evaluation, which is non-longitudinal and does not have a control group, and highlight insights that may inform similar efforts at other institutions. 
Finally, we published our course website, syllabus, slides, homework, and \emph{24-chapter course textbook} online: \url{http://mogu-lab.github.io/cs349-fall-2024}.

\section{Related Work} \label{sec:related-work}

We broadly categorize existing approaches to teaching across the AI spectrum into three lenses: the data-focused lens, the task-focused lens, and the methods-focused lens.
Each of these lenses is complimentary, and is emphasized to differing degrees in every course.
Moreover, each is well-positioned to connect different ethical challenges to technical content (though in practice, \emph{most omit ethics altogether}~\cite{garrett2020more,oliver2021undergraduate,baumer2022integrating,suarez2022ethical}).
In \cref{sec:design}, we then describe our approach to ML pedagogy: the ``framework-focused'' lens.

The \textbf{Data-Focused Lens} (e.g.~\cite{baumer2015data,yavuz2020fostering,donoghue2021teaching}) emphasizes the complexities of working with real-world data, and is well-positioned to connect its technical content to middle-layer ethics.
Courses adopting this lens dedicate substantial time to data wrangling (including cleaning, preprocessing, feature extraction, and imputing missing values), as well as data management, storage, and visualization. 
Ethics instruction in this context typically addresses challenges such as biases introduced by data collection, the use of proxy variables that may encode sensitive attributes, informed consent from data subjects, ensuring privacy and confidentiality, and managing the potential misuse or unintended consequences of data.

Next, the \textbf{Task-Focused Lens} (e.g.~\cite{johnson2023integrating,sherman2023practical}) organizes topics around a taxonomy of methods, placing emphasis on selecting and interpreting ML methods and evaluation metrics appropriate for specific tasks. 
It is therefore positioned to connect technical content with top-layer ethics.
As an example, one common approach is to teach a survey of methods divided into the taxonomy of ``supervised'' and ``unsupervised'' learning.
Ethics instruction within this lens typically focuses on challenges arising from the interplay between data properties and model behavior, such as the ``garbage-in, garbage-out'' problem---wherein biased data leads to biased models---as well as the dangers of spurious correlations leading to poor generalization. 

Finally, the \textbf{Methods-Focused Lens} (e.g.~\cite{hu2020bayesian,dogucu2022current,sestir2023new}) centers the theoretical derivations and properties of methods.
This approach typically opens the black-box methods introduced by \emph{task-focused} pedagogy, but due to the mathematical complexity of the material, typically covers fewer methods, each in significantly more depth.
This type of course often contains little ethics.

In contrast, we propose the \textbf{Framework-Focused Lens} (\cref{sec:design}), which provides students with the \emph{language} and \emph{mathematical formalism} to understand how methods are put together, omitting the complex mathematics of inference.
Specifically, we focus on the probabilistic lens for ML, showing students how to formalize their assumptions into directed graphical models (DGMs).
Using this approach, we forego traditional taxonomies of ML---supervised, unsupervised, semi-supervised---by unifying them under one framework, prompting students to be creative, designing methods that transcend this taxonomy.
In cultivating creativity, we empower students to see how all methods require subjective choices, and using case studies, encourage students to unfold consequences of these assumptions on open-ended ethical questions.
This allows us to connect technical content to base-layer ethics.

\section{Course Design: The Framework-Focused Lens} \label{sec:design}

\begin{figure*}[t]
  \centering
  \includegraphics[width=\linewidth]{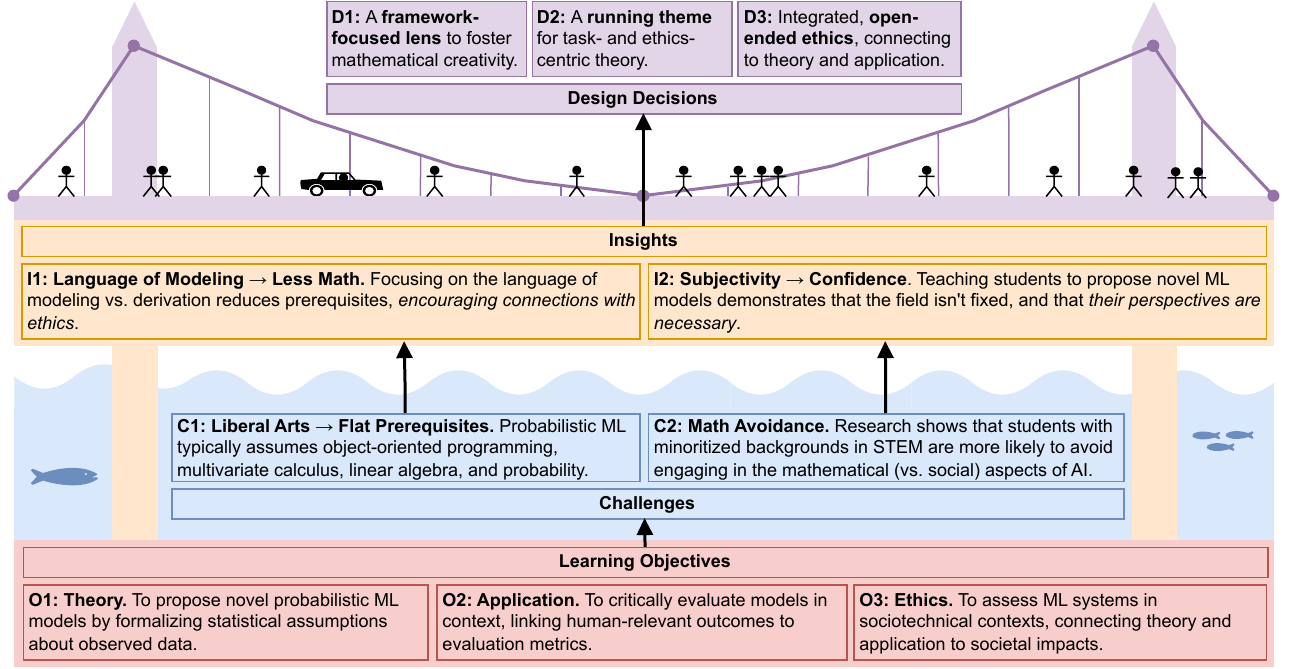}
  \caption{Overview of course design---how our design decisions overcome two key challenges to meet our learning objectives.}
  \label{fig:course-design}
\end{figure*}

We now outline our learning objectives and design decisions, addressing two key challenges that arise in the liberal arts setting.
For an overview, see \cref{fig:course-design}.

\textbf{Learning Objectives.}
Our goal is to empower students to engage in socially and mathematically informed AI discourse.
We designed the course to meet three objectives.
\tagRed{O1 (Theory):} to enable students to propose novel probabilistic ML models by formalizing statistical assumptions about observed data.
\tagRed{O2 (Application):} to teach students to critically evaluate models within the context of their downstream tasks, linking human-relevant outcomes to evaluation metrics.
\tagRed{O3 (Ethics):} to guide students to assess ML systems within their sociotechnical contexts, connecting theory and application to societal impacts, focusing power dynamics.

\textbf{Challenges.}
There are several design challenges we faced in meeting our learning objectives.
\tagBlue{C1: Prerequisites.} 
As a small liberal arts college, our institution intentionally maintains a flat prerequisite structure, making it impractical to require long sequences of preparatory courses. 
In contrast, most probabilistic ML courses are offered at the graduate level, assuming substantial background in object-oriented programming, multivariate calculus, linear algebra, and probability theory. 
We purposefully designed our course to overcome these curricular challenges.
\tagBlue{C2: Math Avoidance.} 
Our institution serves students from minoritized backgrounds in STEM. 
As a result, (i) many students have either lacked access to or have self-filtered out of opportunities to learn advanced mathematics, (ii) our students are more likely to face math anxiety~\cite{lucietto2020math} and to avoid engaging in the mathematical (as opposed to social) dimensions of AI~\cite{barretto2021exploring}. 
In turn, we expect our students to feel inadequate and disengage in AI discourse involving technical jargon, in which their perspectives are desperately needed~\cite{raji2021you}. 
We intentionally designed the course to empower student to participate in both mathematical \emph{and} social types of discourses on AI.

\textbf{Insights.}
We overcome the above challenges based on two insights.
\tagOrange{I1:} \ul{Focusing on mathematical formalism over derivation reduces prerequisites and encourages connections with ethics.}
For ethical discourse grounded in mathematics, we argue it is crucial for students to understand how key assumptions---from model specification to choice of statistical learning algorithm---affect ethics downstream.
Therefore, our course emphasizes both the mathematical formalism and intuition behind model specification, while leveraging a probabilistic programming language (PPL) to ``hide'' the complexities of inference. 
This approach allows us to keep prerequisites minimal, requiring only univariate calculus and experience programming, addressing \tagBlue{C1}.

For instance, in a traditional, methods-focused course, Bayesian linear regression often involves teaching the Bayesian framework followed by a derivation of the posterior, which relies on complicated matrix complete-the-square. 
This derivation obfuscates the Bayesian framework's core principles, the effect of modeling assumptions downstream, and conceptual differences between epistemic and aleatoric uncertainty.
In contrast, our use of a PPL enables approximate posterior inference, freeing students to spend their time exploring model properties empirically, building intuition about Bayesian reasoning, and connecting the mathematics of their assumptions to key ethical dimensions.

\tagOrange{I2:} \ul{Subjectivity fuels creativity, cultivating confidence}~\cite{baker2016underserved}.
To address \tagBlue{C2}, our pedagogy emphasizes the inherent subjectivity of ML methods, repeatedly underscoring how the downstream effects of ML systems depend on crucial human design choices. 
Leveraging this subjectivity, we prompt students to be creative---to propose novel, non-standard ML models for problems they encounter in the course, and to deepen their intuition with experimentation. 
By fostering creativity, we show students that the field is not fixed; rather, their unique perspectives are necessary in ongoing societal discourse around AI.

\textbf{The Framework-Focused Lens.}
We translate our insights into three design decisions, aligned with the 2016 Guidelines for Assessment and Instruction for Statistics Education (GAISE)~\cite{carver2016guidelines}.
\tagPurple{D1:} \ul{A framework-focused lens to foster mathematical creativity.} 
We address \tagBlue{C1}/\tagOrange{I1} by introducing probabilistic model specification and learning as early as possible (``directed graphical models'' or DGMs, and ``frequentist learning'' in \cref{apx-tab:topics}), enabling students to quickly gain a complete picture of the framework.
We do this by initially simplifying the modeling process, focusing only on discrete, fully observed data/models.
Then, we progressively extend their toolkit---e.g.~with continuous probability for regression, the law of total probability for latent variables---with new theory introduced precisely when needed, iteratively reinforcing prior content.
Finally, we use NumPyro~\cite{phan2019composable}---an industry standard, Jax-based~\cite{bradbury2018jax} PPL in Python---to prevent students from getting bogged down by derivations, freeing them to develop intuition (GAISE 2+5~\cite{carver2016guidelines}).

\tagPurple{D2:} \ul{A running theme for task- and ethics-centric theory.} 
We adapt a ``theory through applications''~\cite{nolan1999teaching} approach by creating synthetic datasets that increase in complexity to motivate methodological development, while providing students with a real-like context.
Our running theme tells the story of the ``Intergalactic Hypothetical Hospital'' (IHH), a hospital located in the far reaches of the universe, looking to improve patient care using routinely collected healthcare data.
Throughout the semester, we developed a whimsical lore for the IHH, while providing motivation for methodological development, as well as context for rich, ethical discussions on safety, fairness, and scientific interpretation of ML in healthcare (GAISE 1+3+4~\cite{carver2016guidelines}). 

\tagPurple{D3:} \ul{Integrated, open-ended ethics, connecting to theory and application.}
We designed our ethics sessions to focus on systemic failures rather than individual moral failures, which inappropriately center technologists as ``solitary saviors''~\cite{raji2021you}.
Inspired by~\citet{kirdani2022house}, we paired every technical innovation with counter-narratives that challenge its value (see \cref{apx-tab:topics}).
We chose to instantiate the counter-narratives with real case-studies to cultivate dialectical thinking---to navigate nuanced ethical questions by holding seemingly opposing perspectives in one's head.

\textbf{Homework and Assessment.}
We required students to read relevant chapters in our textbook prior to class, which situate ML methods in the IHH context.
Class met three times a week (twice for 75min and once for 50min). 
Each session consisted of alternating short lectures, pair-work, and class-wide discussions.
Students completed 7 homework assignments, broken into ``checkpoints'' (incomplete submissions, not graded), which allowed the instructors to assess and address common misconceptions in class.
Each homework consisted of problems for individual and group work.
Further, we designed the homework to include open-ended, exploratory problems with ``surprises,'' meant to highlight limitations and important edge cases.
Our course had no exams, and instead motivated students to keep up with the material to better support their peers' learning.
Overall, we found this approach effective.


\section{Evaluation} \label{sec:eval}

We evaluated the course using an anonymous survey, given during class to ensure a high response rate. 
The survey included two types of questions: (a) five-point Likert scale questions asking students to agree/disagree with statements about the course, and (b) open-ended questions about key aspects of the course.
We present the aggregated results from 12 of 16 students, all of whom consented to have their responses shared in aggregate in a publication. 
The full survey is described in \cref{apx:survey}. 

\textbf{Overall.}
As \cref{tab:quantitative} shows, responses to the Likert-scale questions point to the success of the course in achieving our learning objectives. 
The table shows that over $90\%$ of responders agreed with most statements about the course, suggesting they found the theoretical, applied and ethical content enhanced one another, and that they felt more empowered to participate in societal conversations about AI.
Next, we color these responses by summarizing themes from the free-form responses, especially focusing on ethics.

\begin{table}[t]
\caption{Left: statements. Right: percent of students (of 12) that responded with ``somewhat'' or ``strongly agree.''}
\label{tab:quantitative}
\footnotesize
\begin{tabular}{p{7.0cm}r}
\toprule
\textbf{General} & \textbf{\% Agree} \\
\midrule
Overall, I enjoyed the course. & 100.0 \\
I enjoyed the theoretical content of the course. & 100.0 \\
I enjoyed the ethical content of the course. & 91.7 \\
I would recommend the course to my peers. & 100.0 \\[4pt]

\toprule
\textbf{Theoretical Content} & \textbf{\% Agree} \\
\midrule
When new math was introduced, it was clear why it was needed. & 91.7 \\
The derivations in the course materials/lectures enhanced my learning. & 83.3 \\
I feel less intimidated by ML after having taken the course. & 91.7 \\
The probabilistic approach helped me see connections between ML methods. & 91.7 \\
The mathy content enhanced my understanding of the ethics content. & 100.0 \\
The ethics content enhanced my understanding of the mathy content. & 75.0 \\[4pt]

\toprule
\textbf{Applied Content} & \textbf{\% Agree} \\
\midrule
I enjoyed the IHH examples. & 100.0 \\
The IHH examples increased my motivation to learn the material. & 91.7 \\
The IHH examples helped me understand the theoretical content. & 100.0 \\
The IHH examples helped me understand the applied content. & 100.0 \\
The IHH examples helped me understand the ethical content. & 91.7 \\
I related ML topics outside of class to the IHH to help me understand them. & 50.0 \\[4pt]

\toprule
\textbf{Ethical Content: \textit{after} this course, I feel \textit{better prepared} to...} & \textbf{\% Agree} \\
\midrule

Participate in broader societal conversations on AI. & 100.0 \\
Reason about the power dynamics created by ML systems. & 100.0 \\
Reason about ethical questions related to data collection. & 100.0 \\
Reason about how to evaluate ML system in their sociotechnical context. & 100.0 \\
Interrogate the mathematical assumptions behind an ML system. & 91.7 \\
Apply ML in real-world contexts. & 75.0 \\[2pt] 

\bottomrule
\end{tabular}
\end{table}

\textbf{Ethics: Understanding ML in Sociotechnical Contexts.}
Students found that the IHH context helped them center the sociotechnical context to connect the theoretical topics (e.g.~distributional assumptions) to ethics and downstream impact: e.g.,
\begin{quote}
    ``Having the IHH examples integrated into our problem sets [...] helped me to shift my focus from merely understanding and learning the statistical models of ML to critically examining the assumptions I make about the data and its distributions. This [...] allowed me to understand the importance of making assumptions and their potential to cause harm. So, by framing the problems around these intergalactic beings, it encouraged me to recognize the impact and consequences of my work and implementation.''
\end{quote}
The IHH made current research in the intersection of AI and healthcare more accessible to students.
We used this to encourage dialectical ethical thinking, asking students to approach current, open ethical questions with nuanced, conflicting perspectives: e.g.,
\begin{quote}
    ``In my previous ML course, there was some discussion of ethics but it oftentimes felt too spoon-fed, as we were just given an exercise (sometimes optional) and had to arrive at the correct answer [...]. In this course, the discussions about ethics felt more open-ended. We read different articles and research papers that exposed nuances and specific case studies, instead of just telling us a very singular idea. [...] having to arrive at our own conclusions made the ethics aspect of this class more meaningful.''
\end{quote}

\textbf{Ethics: Questioning Narratives of ``Neutrality'' and ``Objectivity'' of ML.}
We approached this in two ways.
First, we emphasized the subjective choices behind each ML method---such as selecting distributions, functional relationships, and statistical dependencies---to highlight that all methods reflect their developers' values. 
Second, by discussing ML's origins in the Eugenics movement, we underscored that these methods were developed by people just like them and are therefore open to critique~\cite{kennedy2024teaching}: e.g.,
\begin{quote}
    ``I was entirely unaware of the history of statistical tools and the influence eugenics had on them. Discovering this history was both surprising and unsettling for me. 
    It was eye-opening to realize how the field of statistics, which I had always viewed as objective and neutral, had been influenced by such controversial and harmful ideologies.''
\end{quote}
By tracing the origins of modern ML to the Eugenics movement, we further prompted students to consider whether ML practice or its theoretical foundations implicitly carry eugenic ideals. 
For instance, we analyze the Maximum Likelihood Estimator (MLE) for regression, and argue how, by minimizing average error, it favors the majority and can produce disparate outcomes across groups.
We then invite students to reflect on their own uncritical acceptance of MLE in the course, and to consider how a similar lack of scrutiny has influenced the development of ML throughout its history.
Finally, using everything they had learned, we ask students to introduce a latent variable into the model and determine whether it mitigates the issue, linking ethical concerns to modeling assumptions.
Many students felt empowered to see how, by re-evaluating their assumptions, they can propose a new approach to regression that overcomes this challenge: e.g.,
\begin{quote}
    ``Our last ethics lessons were especially eye-opening. 
    My favorite activity was drawing out the DGM and deriving the expectation, connecting all the skills we learned previously to the ethics lesson. 
    It demonstrated that we have the potential to transform the field of ML and that the pervasive biases rooted in it don't have to define its future.''
\end{quote}

\textbf{Empowerment: Overcoming Math Avoidance.}
Most of our students entered the course doubting their technical abilities, and all found the course difficult.
There are several themes that emerged from the survey, showing how our course design enabled students to overcome their self-doubt.
First, students appreciated the subjectivity of the modeling process---that their most important choices are their modeling assumptions, which, by definition, can only be obtained through intuition and experimentation: e.g.,
\begin{quote}
    ``I always doubted my place in more `difficult' CS topics like ML and AI. This course delivered ML in an easily digestible way despite how difficult the topics themselves were. We were encouraged to experiment with [...] the models introduced in class, which helped enhance my understanding without making me feel pressured that there was a ``right'' answer all the time.''
\end{quote}
Second, students appreciated our framework-focused lens.
Although mathematically and conceptually challenging, they found this unified perspective showed them how methods are interconnected, giving them confidence they obtained a deeper understanding: e.g.,
\begin{quote}
    ``The other ML courses I've taken were from an optimization approach instead of a probablisitic approach. In those courses, a new framework had to be introduced every time we learned a new method, which made the methods seem very different from one another. The probablistic approach [...] illustrated how the methods are interconnected [...]. This was also the most math-heavy ML course I've taken so far---we talked about into logs, products, integrals, and matrices. I really liked learning about the math and also the implementation because it makes the method seem more concrete and justified. In the other ML courses I've taken, I sometimes feel that the methods we learned are some mystical concept that we should believe in [...].''
\end{quote}
This framework-focused lens further reinforced the idea that the ML process is creative, inviting students as active participants in the development of ML, and not just in its application: e.g.,
\begin{quote}
    ``Coming into this course, I was ready to use popular Python libraries, such as Scikit-learn [...]. In actuality, this course offered more than that. It allowed me to rebuild my foundation in ML, starting with statistical inferences---deciding what distributions to use to represent the data, visualizing relationships using directed graphical models, and mathematically representing those relationships, before reasoning about appropriate inference methods. Beyond solidifying my knowledge, the course also made it flexible by encouraging us to question and critically examine the statistical methods we've used and their implications. The ML process is not rigid and if we want to thoroughly address and mitigate any potential biases and harms, we need to be deliberate with our choices and assumptions, even if that means adding more steps to the process. This course has truly evolved my understanding of ML, and I found every aspect rewarding (with the ethics portion being the most impactful).''
\end{quote}
Third, our framework-focused lens enabled students to understand advanced topics, further increasing their technical confidence: e.g.,
\begin{quote}
    ``I really enjoyed how we slowly developed the way towards more complex models, using a little bit more new math at a time, but still building on top of familiar ideas. A culminating moment for me was when we learned about variational auto-encoders. While there was some new math (like KL divergence), I felt confident learning about this pretty novel technique because I was already comfortable with DGMs [...] the other ML or statistics courses I've taken were more of a survey class, so different methods sometimes felt more disconnected from each other.''
\end{quote}
Finally, students report being able to understand and critique research papers, instead of simply taking them as truth: e.g.,
\begin{quote}
    ``It felt so rewarding to be able to derive and understand the difficult math in this course. Probability and statistics have never been my forte, so it felt great to be able to ground my understanding in CS specifically since it is what I plan to continue doing in the future! It was also rewarding to finally understand the research papers I read outside of class (enough to even question the methods being used).''
\end{quote}

\textbf{Empowerment: Understanding Your Perspective is Needed and Valued.}
In exploring so many ethical issues with ML-based systems, we were worried students would become discouraged from continuing in the field.
We worked hard to communicate with them that open ethical challenges does not mean ``abandon ship;'' rather, it means that we, as a society, don't know what to do, and we need their fresh perspectives to shape the path forward: e.g.,
\begin{quote}
    ``[The ethics sessions] built perfectly upon each other and seriously developed the larger narrative for me. [... They] made me question whether the [AI for healthcare work] I want to do is worth doing, but I quickly realized how such revelations make it even more important for me to pursue these goals after receiving guidance + support from my peers and professors. 
    I actually now feel more empowered as I believe I've gathered the skillset to criticize and question work being done in the industry.''
\end{quote}
\begin{quote}
    ``I've gained confidence in ML, whether it's through the understanding of the mathematical and probabilistic modeling theories, implementing models from the ground up using numpyro, or the deep dive into the ethics of ML that makes me feel like I can have my own unique perspective in the field.''
\end{quote}
We attribute students feeling more like they belong to developing their perspective on ethics: e.g.,
\begin{quote}
    ``I now feel that I have a framework to consider the ethics of not just ML/AI but technical systems in general and ensure that I don't do harm while working in tech, so I feel more comfortable pursuing ML/AI. I also have more self-confidence and feel that I belong in the field.''
\end{quote}
Finally, some students felt empowered to use what they learned in class to further their advocacy goals: e.g.,
\begin{quote}
    ``After this course, I am deeply curious about the uses of ML/AI and the types of models that we could create. I see myself further pursing this field of study to align it with my goal of advocating for marginalized communities with the power of numbers and tools taught in this course.''
\end{quote}

\textbf{Empowerment: Continuing in ML/AI.}
Despite how math-heavy the course was, students still found it accessible: e.g.,
\begin{quote}
    ``Now, since I've realized how accessible ML can be, I am considering furthering my ML education and taking more classes! And ML has definitely become more interesting to me, because I can understand how it works. For example, ``marginalizing out'' a latent variable makes so much sense to me and I can see how it's necessary for our final desired product. It's things like this that keep me encouraged.''
\end{quote}
Students felt confident enough by the end of the course that they considered continuing to pursue ML: e.g.,
\begin{quote}
    ``This class made me [...] feel that I could be able to understand and even work in something like ML. This class has even made me want to pursue research in ML, which I had *never* thought I would have the skillset to do before. I've gained so much more confidence in my ability.''
\end{quote}
And while the course was challenging, students felt like it shaped how they see the field: e.g.,
\begin{quote}
    ``My relationship with AI/ML evolved throughout the semester. I experienced various feelings of conflict, confusion, frustration, etc. It was upsetting to think about how I might be contributing to the negative systemic cycles embedded within our society. I thought I knew ethics, but I realize that I had barely skimmed the surface. I feel that I now will never be able to not think about the ethics of my work. I feel that this course has changed the trajectory of my career as I try to intertwine this new knowledge into my work for positive impact.''
\end{quote}

\section{Discussion, Limitations, and Future Work} \label{sec:discussion}

Our results suggest that the course was well-received, and that students felt it provided them with an accessible, task-focused, and ethics-focused introduction to ML, empowering them to confidently connect theory to applications to ethics.

\textbf{Design: Future Work.}
While by the end, students emerged with a complete modeling toolkit, we would have loved to give them more open-ended modeling problems.
In the future, we plan to split the course in two: the first part will focus on the basic modeling toolkit, leaving generative and Bayesian models to a second, advanced course. 
This will allow more hands-on practice in the first part, making room for an open-ended final project in the second, where students design novel ML models together with researchers on campus.
Next, while the course currently integrates lenses from critical theory into its ethics, this is no substitute to pedagogy by social scientists and humanists~\cite{johnson1994should,raji2021you}. 
We hope to co-teach a truly interdisciplinary version of the course in the future.

\textbf{Evaluation: Future Work.}
We hope to formally evaluate the course via a longitudinal study, comparing its impact on students' long-term ethical engagement with other pedagogical approaches.

\textbf{Two Barriers to Implementation.}
First, the class was designed for a cohort of 18 students, allowing for interactive, personalized instruction, and for building the trust necessary for meaningful ethics discussions and addressing math avoidance.
This may be challenging to scale to larger class sizes.
Second, many instructors may feel unqualified to teach ethics due to limitations in their own training~\cite{johnson1994should}, and staffing constraints can make it difficult to find qualified co-instructors with complimentary expertise.


\begin{acks}
We thank FASPE (Fellowships at Auschwitz for the Study of Professional Ethics), and especially David Goldman, Rebecca Scott, Thorsten Wagner, Maya Dobler, Mary Gray, and Lindsey Cameron, as well as the 2024 Design and Technology Cohort, for discussions about professional ethics that helped shape this course.
We thank members of Wellesley Computer Science Department and Data Science Major for their thoughtful feedback throughout the development of this course.
\end{acks}

\bibliographystyle{ACM-Reference-Format}
\balance
\bibliography{references}

\clearpage
\onecolumn
\appendix

\section{Summary of Topics}

\definecolor{topics}{gray}{0.65}
\definecolor{theory}{gray}{0.75}
\definecolor{coding}{gray}{0.85}
\definecolor{data}{gray}{0.95}
\definecolor{ethics}{gray}{1.0}

\begin{table*}[h]
\caption{Summary of topics, connecting theory to application to ethics. See details in course materials and online textbook: \url{http://mogu-lab.github.io/cs349-fall-2024}.}
\label{apx-tab:topics}

\scriptsize

\begin{tabularx}{\linewidth}{p{0.005\linewidth} | >{\columncolor{topics}}p{0.07\linewidth} | >{\columncolor{theory}}p{0.2\linewidth} | >{\columncolor{coding}}p{0.2\linewidth} | >{\columncolor{data}}p{0.2\linewidth} | >{\columncolor{ethics}}p{0.2\linewidth}}
 & \textbf{TOPIC} & \textbf{THEORY} & \textbf{IMPLEMENTATION} & \textbf{DATA + TASK} & \textbf{ETHICS} \\ \hline \hline
\multirow{4}{\linewidth}{\rotatebox[origin=r]{90}{\textbf{ Directed Graphical Models}}} & Introduction & Conceptually, what is Probabilistic ML? Why use it? & Vectorization in \texttt{jax}. & N/A  & \textbf{Societal Misconceptions about AI:} How they shape participation in the field, and the need for diverse perspectives. \\ \cline{2-6}
 & Probability \newline (Discrete) & Distributions as generative models, and properties of discrete distribution: support, probability mass functions (PMFs), common distributions and their parameters. & Exploratory data analysis (EDA) with \texttt{pandas} and \texttt{matplotlib}. & IHH Emergency Room (ER): EDA of health challenges. & \multirow{2}{\linewidth}{\textbf{The Ethics of Data:} What's lost when representing people as data? Is data ``neutral''? That is, how is data collection and analysis shaped by power dynamics, societally informed research questions? How do we responsibly collect and use data?} \\ \cline{2-5}
 & Conditional \newline Probability \newline (Discrete) & Conditional distributions as predictive models, PMFs with parameters as functions. & Distributions in \texttt{numpyro}, reproducibility with splittable random keys in \texttt{jax}. & IHH ER: Predicting health challenge given day of week, predicting likelihood of hospitalization and use of antibiotics. \\ \cline{2-5}
 & Joint \newline Probability \newline (Discrete) & Joint distributions as multivariate generative models, factorizing joint distributions, denoting statistical dependencies via DGMs, ancestral sampling. & Ancestral sampling and computing joint log-probabilities via \texttt{numpyro} distributions. & IHH ER: Identifying ``unlikely'' patients with generative models. \\ \hline
\multirow{3}{\linewidth}{\rotatebox[origin=r]{90}{\textbf{ Frequentist Learning}}} & Maximum \newline Likelihood \newline Estimation \newline (MLE) & IID sampling and plate notation for DGMs, writing MLE formal problem statement for different DGMs, theoretical properties of MLE (briefly). & Translating DGMs to \texttt{numpyro} models via primitives (e.g.~\texttt{param}, \texttt{sample}), fitting models to data and interpreting parameters. & (Same as for Joint Probability) & \multirow{3}{\linewidth}{\textbf{The Ethics of Learning from Data:} The value/myth of generalizability in ML for healthcare, the ethics of statistical significance testing, the ``neutrality'' of ML model specification and learning, and what is lost when translating model's learned parameters to human-understandable ``knowledge''?} \\ \cline{2-5}
 & Optimization & What is optimization, advantages of analytic vs. numeric optimization, the gradient descent algorithm, local and global optima. & Diagnosing training issues by matching plots of loss vs. epoch to hypothetical shapes of loss landscapes and learning rates. & (Same as for Joint Probability) \\ \cline{2-5}
 & Probability \newline (Continuous) & Revisit DGMs, now with continuous distribution, properties of cumulative and probability distribution functions, common distributions and their parameters, expectations. & Implementing discrete-continuous DGMs and fitting them with MLE in \texttt{numpyro}. & IHH’s Center for Telekinesis Research (CTR): How do physiological conditions (discrete) affect telekinetic ability (continuous)? \\ \hline
\multirow{4}{\linewidth}{\rotatebox[origin=r]{90}{\textbf{ Predictive Models}}} & Regression & DGM and MLE for predictive models, instantiation regression, correlation vs. causation, connection with mean square error (MSE) minimization. & Implementing linear and polynomial regression in \texttt{numpyro}. & IHH’s CTR: Understanding the relationship between ``glow'' and ``telekinetic ability.'' & \multirow{4}{\linewidth}{\textbf{The Ethics of Predictive Models in Sociotechnical Systems:} Conducting a broader impact analysis for case studies in the intersection of AI and healthcare---identifying users, stakeholders, affected communities, etc. as well as their power dynamics, responsibility, and more.} \\ \cline{2-5}
 & Classification & Instantiation classification as a specific predictive model, generalized predictive models. & Implementing linear and non-linear classification in \texttt{numpyro}. & IHH's CTR: Understanding the relationship between age, the dosage of medication, and telekinetic control. \\ \cline{2-5}
 & Neural \newline Networks (NNs) & Shortcomings of polynomial bases, expressivity through function composition, matrix representation of NNs, open challenges with NNs. & Implementing NN regression in \texttt{numpyro}, experimenting with architecture and optimization, matching inductive bias of different regression models to task. & (Same as for Regression) \\ \cline{2-5}
 & Model \newline Selection \newline and \newline Evaluation & Probabilistic view of over/under-fitting, log-likelihood, metrics for regression and classification and their shortcomings, train/validation/test splits. & Intuition matching model fit to metric, intuition for need to use validation sets, empirically discovering and diagnosing shortcomings of regression/classification metrics. & IHH’s Center for Rare Disorders: For regression, reducing pain from Antenna Inflammation with ``space beam''---how much is too much? And understanding disparities in data consisting of multiple populations. For classification, predicting rare occurrences of Antenna Inflammation. \\ \hline
\multirow{2}{\linewidth}{\rotatebox[origin=r]{90}{\textbf{ Generative Models}}} & Gaussian \newline Mixture \newline Models (GMMs) & Latent Variables, Law of Total Probability, DGM and MLE for GMMs. & Fitting GMMs to data, diagnosing optimization challenges and cluster non-identifiability. & (Same as for Continuous Probability) & \multirow{2}{\linewidth}{\textbf{The Ethics of Generative Models in Sociotechnical Systems:} Conducting a broader impact analysis (like above) for real case studies of harms caused by large language and text-to-image stable diffusion. Discuss new approaches that use qualitative (instead of only quantitative) methods for identifying and quantifying harm in complex, black-box models.} \\ \cline{2-5}
 & Factor \newline Analysis (FA) & The non-linear FA model, computational challenges with MLE due to marginalization, Monte-Carlo integration. & Implement neural FA in \texttt{numpyro}, explore data generated by interpolation in the low-dimensional latent space. & IHH’s Center for Epidemiology: Learning the distribution of microscope images of intergalactic viruses to quantify disease variability. \\ \hline
\multirow{2}{\linewidth}{\rotatebox[origin=r]{90}{\textbf{ Bayesian Models}}} & Priors \newline and \newline Posteriors & What are aleatoric and epistemic uncertainty and why we need them, Bayesian specification (priors and likelihoods) and DGM representation, Bayes' rule for posterior inference. & \multirow{2}{\linewidth}{Implementing Bayesian linear, polynomial, and NN regression in \texttt{numpyro}, gain intuition for uncertainty via model comparison.} & \multirow{2}{\linewidth}{IHH’s CTR: Understanding the relationship between age and telekinetic ability---what to do for out-of-distribution patients?} & \multirow{2}{\linewidth}{\textbf{The Ethics of Uncertainty and Interpretability in Human-AI Systems:} What do we need for effective human-AI collaboration? What are current barriers? For example, can perceptions of fairness be manipulated with explainable AI (XAI)? How do power dynamics and responsibility shift by introducing XAI?} \\ \cline{2-3}
 & Posterior \newline Predictives & Bayesian model averaging, DGM for jointly representing train and test data, laws of conditional independence, derivation of posterior predictive for an array of popular models. &  &  \\ \hline
\multirow{3}{\linewidth}{\rotatebox[origin=r]{90}{\textbf{ Synthesis}}} & Variational Autoencoders & Intractability of expectations, importance sampling, Jensen's inequality, the variational evidence lower bound, reparameterization trick. & N/A & (Same as Factor Analysis) & \multirow{2}{\linewidth}{\textbf{The Ethics of ML---A View from History:} What was the Eugenics movement? How did it shape the birth of statistics and modern ML? How do Eugenics ideals manifest in the practice of ML and in its underlying math? How do we reckon with this history of ML?} \\ \cline{2-5}
& Special Topics & Examples include: Gaussian Processes, dynamical systems, the ethics of AI in the criminal justice system, guest lectures, etc. & & \\
\bottomrule
\end{tabularx}
\end{table*}

\FloatBarrier
\newpage
\twocolumn

\section{Evaluation Survey}

\subsection{Survey Questions} \label{apx:survey}

\noindent \textbf{Preamble.}
The purpose of this survey is to: (1) help improve future iterations of the course, and (2) to inform a paper describing the course's approach to ML pedagogy, in case faculty at other institutions want to adopt it.
Your thoughtful engagement with this survey is super important to us.

\noindent \textbf{Questions.}
\begin{enumerate}
    \item I give the instructor consent to use the anonymously collected data from this survey in a future publication describing their approach to ML pedagogy. Data from this survey may be presented in aggregate, after having been anonymized, so that no identifiable information can be extracted. [Yes/No]
    
    \item The following items ask about your experiences working with data from the Intergalactic Hypothetical Hospital (IHH). Using the scale below, rate the extent to which you agree or disagree with each statement: [5-point likert scale]
    \begin{enumerate}
        \item I enjoyed the IHH examples.
        \item The IHH examples increased my motivation to learn the material.
        \item The IHH examples helped me understand the theoretical content of the course.
        \item The IHH examples helped me understand the applied content of the course.
        \item The IHH examples helped me understand the ethical content of the course.
        \item I relate ML topics I encounter outside of class to the IHH to help me understand them.
    \end{enumerate}
    
    \item If you found the IHH examples enhanced your learning, in what ways did they? If you found they hindered your learning, in what ways did they? Please be specific with concrete examples from the course. [open response]
    
    \item If you've taken other ML or statistical modeling course(s), in what ways was the applied content of this course different (i.e. application to IHH data)? How did each difference affect your learning? Please provide specific examples. [open response]
    
    \item Ethics: After having taken this course, I feel better prepared to... [5-point likert scale]
    \begin{enumerate}
        \item Participate in broader societal conversations on AI.
        \item Reason about the power dynamics created by ML systems.
        \item Reason about ethical questions related to data collection.
        \item Reason about how to evaluate ML system in their sociotechnical context.
        \item Interrogate the mathematical assumptions behind an ML system.
        \item Apply ML in real-world contexts.
    \end{enumerate}

    \item If you've taken other ML or statistical modeling course(s), in what ways was the ethics content of this course different? How did each difference affect your learning? Please provide specific examples. [open response]

    \item Using the scale below, rate the extent to which you agree or disagree with each statement: [5-point likert scale]
    \begin{enumerate}
        \item When new math was introduced, it was clear why it was needed.
        \item The derivations in the course materials and lecture enhanced my learning.
        \item I feel less intimidated by ML after having taken the course.
        \item I found the probabilistic approach to ML helped me see connections between different ML methods.
        \item The mathy content enhanced my understanding of the ethics content.
        \item The ethics content enhanced my understanding of the mathy content.
    \end{enumerate}

    \item If you've taken other ML or statistical modeling course(s), in what ways was the theoretical content of this course different? How did each difference affect your learning? Please provide specific examples. [open respnose]

    \item Using the scale below, rate the extent to which you agree or disagree with each statement: [5-point likert scale]
    \begin{enumerate}
        \item Overall, I enjoyed the course.
        \item I enjoyed the theoretical content of the course.
        \item I enjoyed the ethical content of the course.
        \item I would recommend the course to my peers.
    \end{enumerate}

    \item Overall, what did you find rewarding about the course? [open response]

    \item Overall, what did you find challenging about the course? [open response]

    \item Before to having taken the course, what were your opinions of ML/AI? [open response]

    \item After having taken the course, did your opinion(s) change? If so, in what way(s)? Please provide specific examples. [open response]

    \item Before to having taken the course, what was your relationship with ML/AI? (e.g.~how did you feel about studying/pursuing it?). [open response]

    \item After having taken the course, did your relationship change? If so, in what way(s)? Please provide specific examples. [open response]

    \item Is there anything else you'd like us to know about your experience in the course? [open response]
\end{enumerate}

\subsection{Survey Results}

\cref{apx-fig:likert} below presents the full results from \cref{tab:quantitative}.

\FloatBarrier

\begin{figure*}[p]
  \centering
  \includegraphics[width=\linewidth]{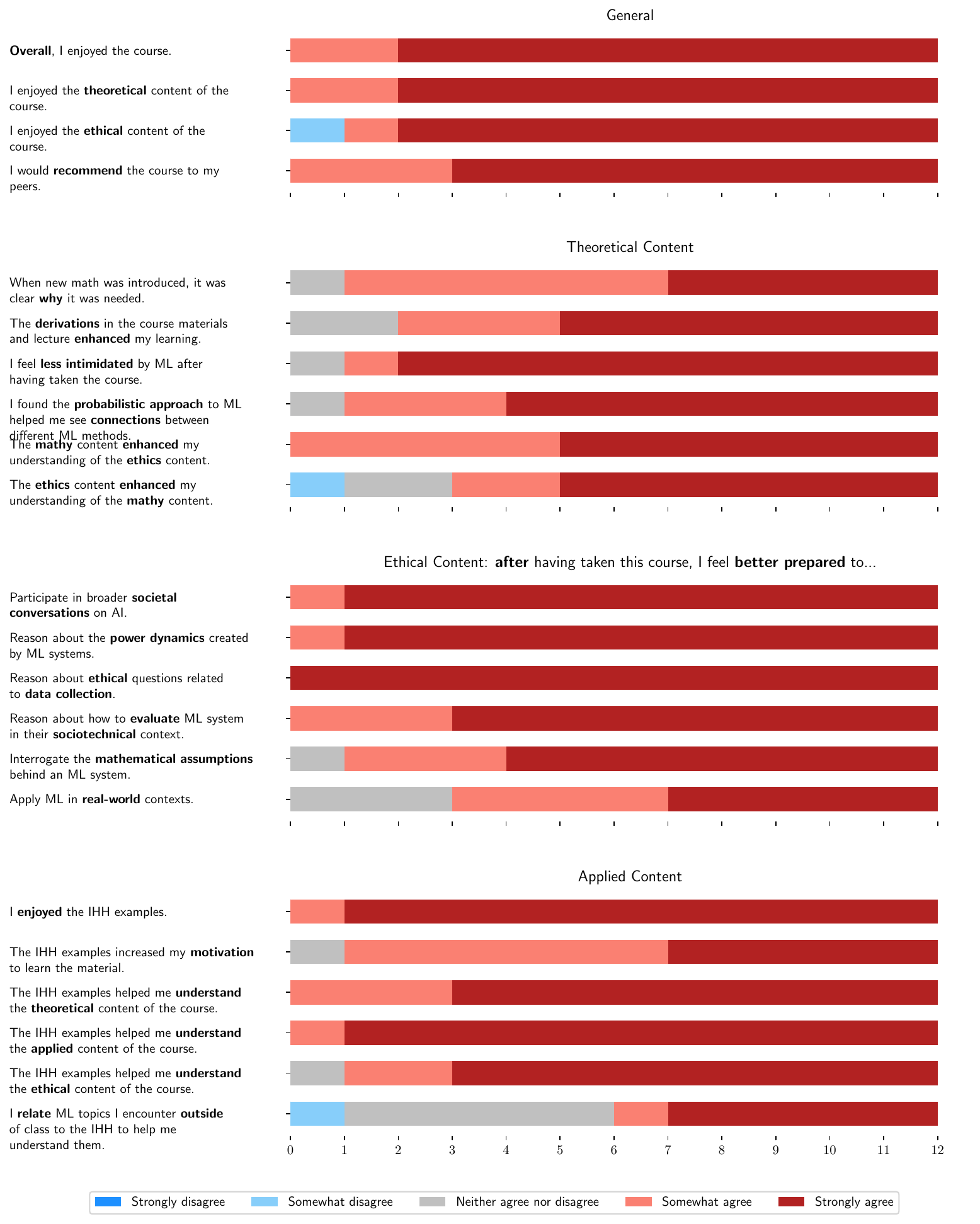}
  \caption{Responses to the likert-scale survey questions; students responded positively to the design and delivery of the course.}
  \label{apx-fig:likert}
\end{figure*}


\end{document}